The original manuscript was finished in 2017 but not submitted in time due to the delay in the preparation for source code release. Though some of the references and assumptions in this manuscript are dated, the key messages which the authors were trying to convey remain valid today amid the continuous advancement of one-/two-dimensional materials:

a) The intrinsic performance gain through the adoption of two-dimensional materials as the FET channel materials can be greatly limited by the parasitic resistances.
b) Design technology co-optimization across the boundaries between devices, interconnects, circuits, and systems can help direct resources toward the most promising candidates and maximize the values of new technologies.
c) Machine learning algorithms can facilitate holistic design technology co-optimization.

As a result, we decide to still upload the manuscript to arXiv now and hope that our findings and insights can still benefit the research community.

# Device-to-System Performance Evaluation: from Transistor/Interconnect Modeling to VLSI Physical Design and Neural-Network Predictor

Chi-Shuen Lee, Brian Cline, Saurabh Sinha, Greg Yeric, and H.-S. Philip Wong, *Fellow*, IEEE

*Abstract*—We present a DevIce-to-System Performance EvaLuation (DISPEL) workflow that integrates transistor and interconnect modeling, parasitic extraction, standard cell library characterization, logic synthesis, cell placement and routing, and timing analysis to evaluate system-level performance of new CMOS technologies. As the impact of parasitic resistances and capacitances continues to increase with dimensional downscaling, component-level optimization alone becomes insufficient, calling for a holistic assessment and optimization methodology across the boundaries between devices, interconnects, circuits, and systems. The physical implementation flow in DISPEL enables realistic analysis of complex wires and vias in VLSI systems and their impact on the chip power, speed, and area, which simple circuit simulations cannot capture. To demonstrate the use of DISPEL, a 32-bit commercial processor core is implemented using theoretical n-type $MoS_2$ and p-type Black Phosphorous (BP) planar FETs at a projected 5-nm node, and the performance is benchmarked against Si FinFETs. While the superior gate control of the $MoS_2$/BP FETs can theoretically provide 51% reduction in the iso-frequency energy consumption, the actual performance can be greatly limited by the source/drain contact resistances. With the large amount of data generated by DISPEL, a neural-network is trained to predict the key performance metrics of the 32-bit processor core using the characteristics of transistors and interconnects as the input features without the need to go through the time-consuming physical implementation flow. The machine learning algorithms show great potentials as a means for evaluation and optimization of new CMOS technologies and identifying the most significant technology design parameters.

## I. INTRODUCTION

As conventional scaling of Si transistors and Cu interconnect began to face significant difficulties [1-3], candidates to complement Si and Cu to extend CMOS technology scaling in the sub-10-nm technology nodes are researched extensively. Active development for new technologies includes nanowires [4], two-dimensional (2D) semiconductors [5], and carbon nanotubes (CNT) for transistors [6]; and cobalt, graphene, and CNT for interconnects[1] [7-8]. The high cost of developing a new technology makes it vital to gain an early understanding of their potential benefits. Early insights into the key performance detractors help focus development efforts such that resources can be directed toward the most important challenges. However, accurate technology assessment at early stage is difficult due to growing impact of parasitic and interconnect resistances and capacitances, which depend on the cell layouts and system architecture. Traditional simple benchmarks [9-10] become insufficient to capture the complexity of interconnects in a Very-Large-Scale Integration (VLSI) system. Ring oscillators with fixed fan-out and wire loads is a common benchmark to compare power and speed of different technologies, whose wire lengths are typically estimated by the average wire length of VLSI circuit implementations or a multiple of the contacted gate pitch (CGP). While this simple approach provides some insights, estimation of the wire load is not easy because different transistor drive current and capacitance can lead to different optimal circuit topologies (see Section IV). Other system-level models [11-14] employed Rent's rule [15] to derive a stochastic wire distribution, optimize the interconnects given a gate delay model, and predict the final chip power, speed, and area. Despite the high-level abstraction of architectures and wiring optimization, many empirical parameters must be decided (e.g. the Rent's constants), which can be architecture and/or technology dependent and need careful calibration. Recent works resorted to full physical design flows to assess system-level performance by actual implementation of a VLSI system or a large circuit module [16-19]. This approach provides the most accurate performance evaluation as complex wiring is considered realistically. Therefore, in this paper we present an end-to-end DevIce-to-System Performance EvaLuation (DISPEL) workflow that automates both Process Design Kit (PDK) development and physical design flows, which enables efficient system-level performance evaluation of transistor and interconnect technologies. Similar method has been employed in [19] to evaluate the benefits of carbon nanotube field-effect transistors (FETs). Here the emphasis is on: (i) the methodology of performance evaluation, (ii) impact of parasitics on performance, and (iii) applying neural-network (NN) models to efficient processor core-level performance evaluation.

C.-S. Lee was with the Department of Electrical Engineering, Stanford University, Stanford, CA 94305 USA. He is now with Google LLC. (e-mail: chishuen@alumni.stanford.edu).
Brian Cline is with ARM Ltd.
Saurabh Sinha was with ARM Ltd. He is now with Apple Inc.
Greg Yeric was with ARM Ltd. He is now with Cerfe Labs Inc.
H.-S. P. Wong is with the Department of Electrical Engineering, Stanford University, Stanford, CA 94305 USA (e-mail: hspwong@stanford.edu).

---

[1] The term "interconnect" in this paper refer to both wires and vias.

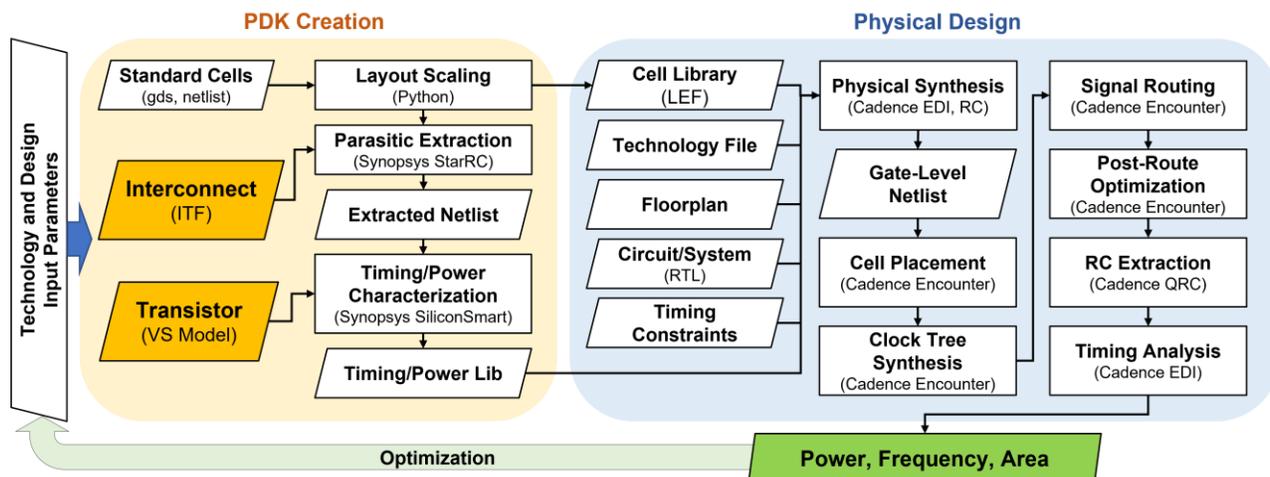

Fig. 1. Overview of the DevIce-to-System Performance EvaLuation Platform (DISPEL) workflow. Corresponding EDA tools or file formats are specified in the parentheses. Acronyms–LEF: Library Exchange Format, Lib: Liberty model, RTL: Register Transfer Language, ITF: Interconnect Technology Format, VS: Virtual Source [21].

This paper is organized as follows: the DISPEL workflow is introduced in Section II; the methodology of technology evaluation and optimization is presented in Section III. To demonstrate the use of DISPEL, CMOS devices composed of n-type $MoS_2$ and p-type Black Phosphorous (BP) FETs are used to implement a 32-bit commercial processor core at a projected 5-nm node. The performance is compared against Si FinFETs to stress the impact of device parasitics; in Section IV, NN models are trained to predict the core energy consumption, delay, and die area using the data generated by DISPEL to address the issue of long runtime of physical design flows; in Section V, the NN models are analyzed to understand what has been learnt from the data; finally, in Section VI, we discuss the limitations of DISPEL and the NN models, as well as potential future research directions.

## II. DISPEL: DEVICE-TO-SYSTEM PERFORMANCE EVALUATION PLATFORM

An overview of DISPEL workflow is depicted in Fig. 1. The flow consists of three parts: (i) transistor and interconnect modeling, (ii) PDK development, and (iii) physical design flows. The major steps are described in this section to provide essential knowledge for the rest of the sections.

### A. Transistor and Interconnect Models

The compact Virtual-Source (VS) model [21] is used as a basis for the transistor modeling because the model parameters can be extracted from experimental data and have physical meaning in the quasi-ballistic transport regime. To ensure physically meaningful results, the apparent carrier mobility ($\mu$) [22] and injection velocity ($v$) of VS models are extracted from either experimental data or physical simulations. In Fig. 2, VS models are fitted to the simulated current-voltage (I-V) curves of monolayer n-type $MoS_2$ and p-type BP FETs based on Non-Equilibrium Green's function formalism [23-24] to extract $\mu$ and $v$ (a parameter extractor is provided in [25]). Here $\mu$ and $v$ should be viewed as fitting parameters reflecting the theoretical performance of $MoS_2$ and BP FETs with clear physical meanings [22].

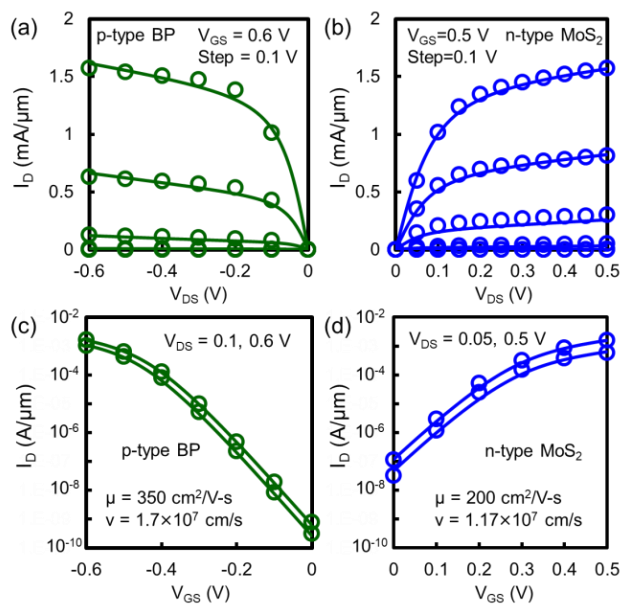

Fig. 2. Virtual Source model (lines) fitted to NEGF-based numerically simulated monolayer n-type $MoS_2$ [23] and p-type black phosphorous (BP) [24] FETs (circles). (a) $I_D$-$V_{DS}$ for BP and (b) $MoS_2$. (c) $I_D$-$V_{GS}$ for BP and (d) $MoS_2$. $L_{GATE}$ = 10 nm and equivalent oxide thickness = 0.7 nm for both FETs.

Characteristics of interconnect wires, vias, and interlayer dielectrics are specified in an Interconnect Technology Format (ITF) file, including the dimensions, resistivities, and dielectric constants of each layer, which are parameterized for easy experimentation with different interconnect properties. The interconnects can be conceptually categorized into two parts: local middle-end-of-line (MEOL) and intermediate/global back-end-of-line (BEOL) interconnects as illustrated in Fig. 3 (a) and (b). At advanced technology nodes, cell-level parasitics dominate over transistor resistance and capacitance (RC) [26]. One of the dominant components is the metal-to-semiconductor source/drain (S/D) contact resistance ($R_{CON}$). Here $R_{CON}$ is modeled by $\rho_{CON}/(L_{CON}W)$ assuming $L_{CON}$ is much shorter than the transfer length [27], where $\rho_{CON}$ is the specific contact resistivity, $L_{CON}$ is the contact length, and $W$ is the device width. $R_{CON}$ is included in the MEOL parasitic RC during the parasitic extraction step in the workflow rather than in the transistor model because $R_{CON}$ depends on the circuit topology. For instance, when a S/D contact is shared by two parallel transistors, the impact of $R_{CON}$ is amplified, which is an effect difficult to be captured inside a compact model. The BEOL interconnects include Metal-1 (M1) layer and the layers above; copper (Cu) is used as the default materials for wires and vias. Here the Steinhögl model [28] calibrated to experimental data [29] is employed to account for the dependence of Cu resistivity ($\rho_{Cu}$) on the cross-sectional area as shown in Fig. 4. $\rho_{Cu}$ increases dramatically with dimensional scaling due to electron scattering at the surfaces and grain boundaries.

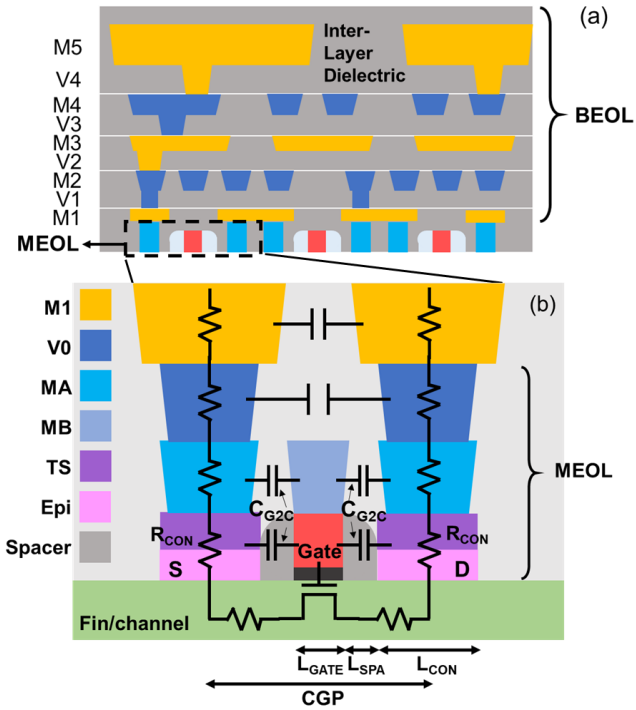

Fig. 3. Illustration of (a) BEOL interconnects and (b) MEOL interconnects and cell-level parasitic RC. MA, MB are local interconnect metals; TS is trench silicide (for Si) or the metals directly connecting to the source (S) and drain (D); $R_{CON}$ is contact resistance; $C_{G2C}$ is gate-to-contact coupling capacitance.

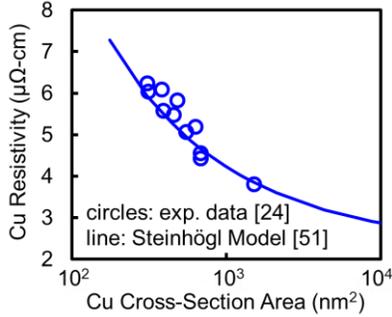

Fig. 4. The Steinhögl model [28] fitted to experimental data [29], used in DISPEL to capture the copper resistivity vs. cross-section area relationship.

### B. *Process Design Kit (PDK) Creation*

The PDK creation flow starts with dimensional scaling of standard cell (SC) layouts. The SC library used in this paper includes more than 100 cells. Fig. 5 illustrates the scaling of an INV_X1 cell (i.e. inverter with 1 finger) for example. The cell width is proportional to CGP and the height is proportional to M2 pitch. All the widths and lengths of the polygons in the layouts are parameterized such that they can be scaled by computer programs (e.g. Python scripts) easily. The Design Rule Checks (DRC) are relaxed to an extent so that layouts can pass the checks as long as two polygons on the same layer do not touch or overlap each other. The relaxed DRCs help us focus on the impact of scaling on performance without worrying about the constraints of photolithography. Then the scaled layouts are tested by Layout Versus Schematic checks to ensure consistency after scaling. With the scaled layouts, netlist templates, and an ITF file, cell-level parasitic extraction is performed by Synopsys StarRC [30] to generate extracted SC netlists with all the parasitic RCs. Timing and power characteristics of SCs in the library are then characterized by Synopsys SiliconSmart [31].

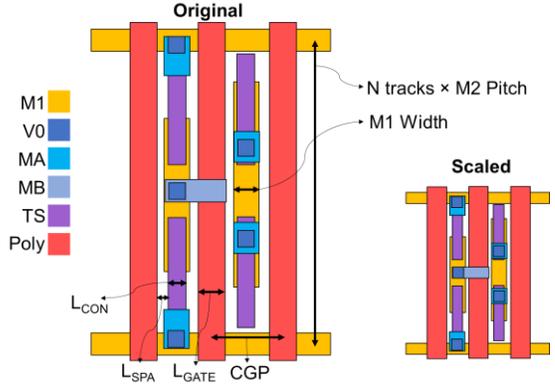

Fig. 5. Illustration of layout scaling of the INV_X1 standard cell (i.e. inverter with 1 finger). Cell height is proportional to M2 pitch where N is the number of tracks, a constant number.

### C. *Physical Design*

A simplified yet complete physical design flow is used to map a design in Register Transfer Language (RTL) into physical geometric representations of all the layers that can be manufactured. The flow involves floor-planning, logic synthesis, Design-For-Test scan chain insertion, cell placement, clock tree synthesis, signal routing, RC extraction, setup-/hold-time violation fix, and static timing analysis. Physical verification and signal integrity analysis are skipped as the focus here is performance. Inputs to the physical design flow include: (i) a technology file defining the metal layers used for signal routing (e.g. minimum metal width and separation); (ii) a circuit/system RTL design to be implemented (e.g. a microprocessor core); (iii) a floorplan specifying the die size, locations of the input/output (I/O) pins, positions of the module macros and blockages; (iv) a design constraint file (e.g. set clock uncertainty, maximum signal transition, etc.); (v) a target clock frequency ($f_{TAR}$). Outputs at the end of the flow are the total chip power consumption, total cell area, and critical-path timing analysis. Rather than employing complex techniques to achieve timing closure after the physical design, an *achieved clock frequency* ($f_{ACH}$), defined as $1/f_{ACH} = (1/f_{TAR}) - t_{SLACK}$, is used as the metric of speed, where $t_{SLACK}$ is the timing slack of the critical path.

TABLE I. PROJECTED 5-NM NODE METAL WIDTHS AND SPACINGS

| Layer | Min Width | Min Spacing |
|---|---|---|
| M1, M2, M3 | 12 nm | 12 nm |
| M4, M5 | 18 nm | 18 nm |
| M6 | 24 nm | 24 nm |
| V1, V2, V3 | 12 nm | 12 nm |
| V4, V5 | 18 nm | 18 nm |

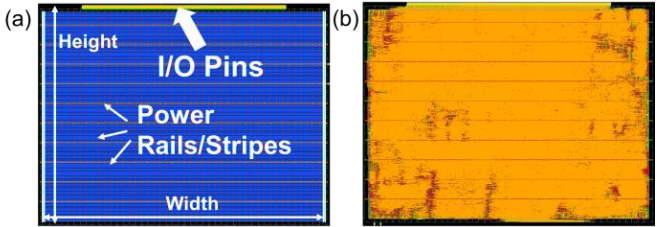

Fig. 6. (a) Floorplan. I/O pins are located on the top of the core to connect to the memory system. The die size may change but the aspect ratio (Height/Width) and I/O locations remain fixed. (b) Top-down view of output GDS layout of an implemented core (the lines in orange/dark red are metal lines).

## III. SYSTEM-LEVEL PERFORMANCE ASSESSMENT OF NEW TRANSISTOR/INTERCONNECT TECHNOLOGIES

A commercial 32-bit microprocessor core is implemented at a projected 5-nm technology node using the monolayer n-type $MoS_2$ and p-type BP FETs (see Fig. 2) to demonstrate the use of DISPEL. The 5-nm node is assumed to have a CGP of 36 nm and a M2 pitch of 24 nm, extrapolated from the 7-nm node [32]. Table I summarizes the minimum widths and spacings of metal wires and vias at different levels. M2 and the layers above are used for signal routing up to M6 as the core is a power-efficient design and relatively simple. $MoS_2$ and BP are selected to represent the 2D layered material family because $MoS_2$ is the most mature among the family [33] and BP has the highest mobility in theory [34]. Evaluation of other transistor technologies has been studied in [16, 35]. This paper focuses on the methodology for performance evaluation, device structure optimization, and analysis of interconnects.

### A. Methodology for Core Performance Evaluation

The processor core implementation flow is simplified as follows: (i) process-voltage-temperature (PVT) variations are not considered, i.e. only the nominal-case transistor and interconnect models at the typical corners are used in the timing and power analysis; (ii) Static Random-Access Memories (SRAMs) are removed from the core because predictions of SRAM performance heavily rely on many other factors such as process variability, which requires separate optimization [36-37] and is beyond the scope of this paper; (iii) a simple floorplan is used for all the core implementations throughout this paper as shown in Fig. 6. All the I/O pins are located on one side of the core to connect to the memory system. While floor-planning can significantly affect the core performance [38], this simple floorplan allows us to focus on the impact of CMOS technologies on the system-level performance to gain early insights. The final outputs of the workflow are core energy consumption, $f_{ACH}$, and total cell area. Throughout this paper these numbers are normalized to an arbitrary number to emphasize the trend rather than the absolute values.

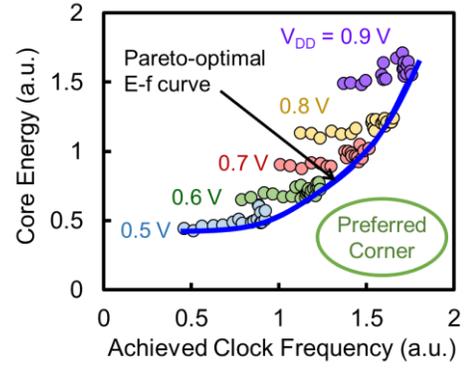

Fig. 7. Pareto-optimal energy-frequency curve for the 32-bit processor core implemented using n-$MoS_2$/p-BP FETs following the methodology in Section III. Each dot represents an implementation for a target clock frequency.

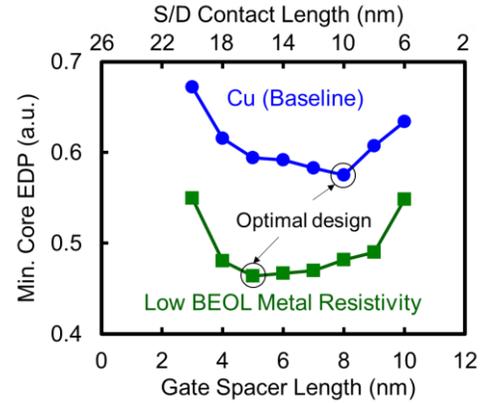

Fig. 8. Minimum core energy-delay product vs. gate spacer length (see Fig. 3b) with fixed CGP = 36 nm and $L_{GATE}$ = 10 nm. The low resistivity case (green) is assumed to have a 0.1× of the BEOL metal resistivity of the Cu case.

The primary performance metric used in this section is the Pareto-optimal energy vs. $f_{ACH}$ trade-off curve (E-f curve), where the energy is calculated by the total power consumption divided by $f_{ACH}$. The E-f curves are generated as follows:
1. For a given supply voltage ($V_{DD}$), tune the threshold voltage in the VS model to meet the target transistor leakage current ($I_{OFF}$ = 1 nA/μm throughout this paper).
2. Scale the standard cell layouts based on the input dimensional parameters (e.g. $L_{GATE}$, $L_{CON}$) and run parasitic extraction to generate extracted netlists.
3. Characterize the timing and power of the SC library.
4. Run physical design flow for a given $f_{TAR}$ to generate the core energy consumption and $f_{ACH}$.
5. Repeat Step 4 over a range of $f_{TAR}$ with a coarse frequency interval (e.g. from 1 to 3 GHz with a step size of 0.2 GHz).
6. Pick the design at the maximum $f_{ACH}$ from Step 5 and scale the die width and height proportionally with a fixed aspect ratio to achieve a cell utilization of 60%.
7. Repeat Step 4 around the maximum $f_{ACH}$ with a finer step size (e.g. a step size of 0.02 GHz).
8. Repeat Step 1 and 3-7 for different $V_{DD}$'s (e.g. from 0.5 to 0.9 V with a step size of 0.1 V).
9. Generate a Pareto-optimal E-f curve by connecting the optimal design points, i.e. the point with the lowest energy consumption for any given $f_{ACH}$, as illustrated in Fig. 7.

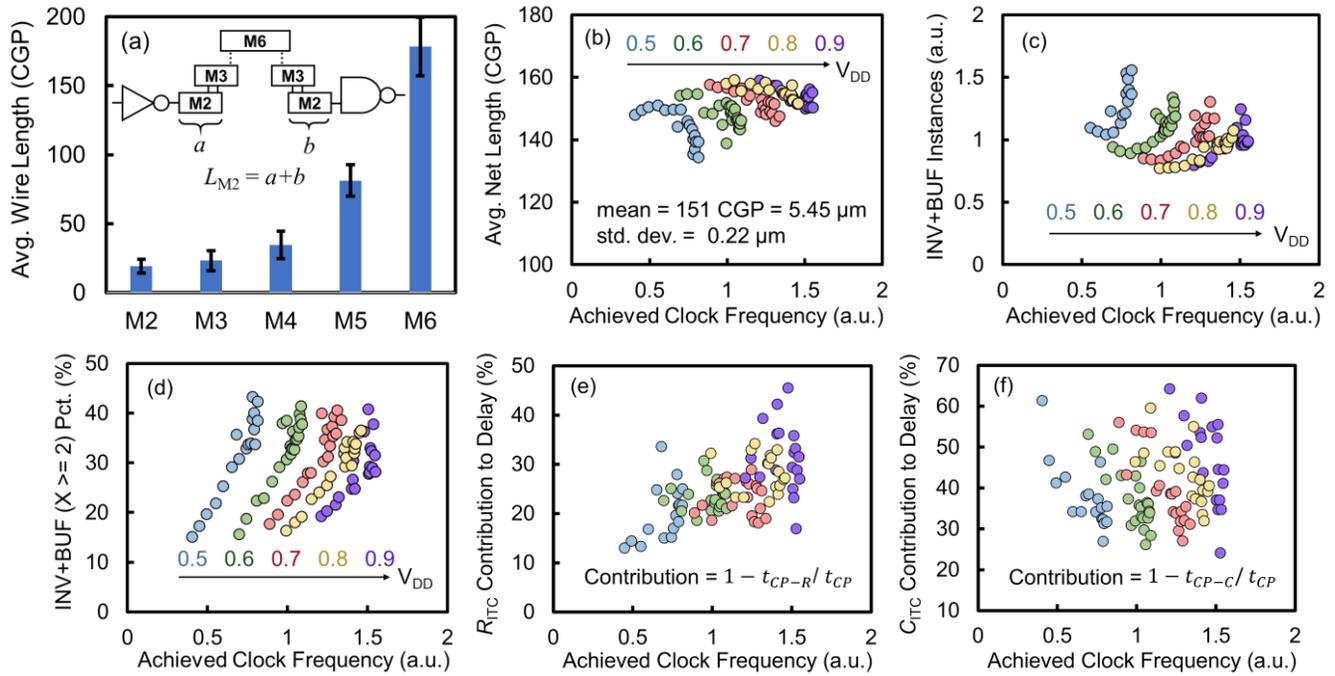

Fig. 9. Statistics of 98 core implementations for different $V_{DD}$'s after timing closure. (a) Average wire length of each metal layer (normalized to CGP = 36 nm) between two connected standard cells on the top 20 critical paths (illustrated by the inset). (b) Average net length over a core. (c) Number of inverter and buffer instances. (d) Ratio of inverter and buffer instances that have two or more fingers to the total number of inverters and buffers. (e)(f) Contributions of interconnect (wires + vias) resistances ($R_{ITC}$) and capacitances ($C_{ITC}$) to the critical path delays ($t_{CP}$), where $t_{CP-R(C)}$ is the delay with $R_{ITC}$ ($C_{ITC}$) zeroed out.

### B. *Device Structure Optimization*

Ultrathin (~1 nm) channel materials (e.g. 1D carbon nanotubes, 2D $MoS_2$) can provide superior electrostatic gate control and carrier mobility to bulk materials (e.g. Si, Ge) [16] at similar channel thickness. To maintain good gate control, the channel body thickness of bulk materials must be thinned down, leading to drastic degradation in the carrier mobility [39]. Good gate control allows further reduction of the gate length ($L_{GATE}$) without increasing subthreshold leakage current; shorter $L_{GATE}$ also helps reduce intrinsic gate capacitance and gives more space to the gate spacer and S/D contacts to lower the cell-level parasitic RC. Trade-offs exist between the lengths of gate spacer ($L_{SPA}$), S/D contacts ($L_{CON}$) and $L_{GATE}$ for a fixed CGP —as $L_{SPA}$ increases, $L_{CON}$ decreases accordingly, resulting in decreasing gate-to-contact coupling capacitance ($C_{G2C}$) but increasing $R_{CON}$. To fully understand the potential of a transistor, it is important to explore the full design space and find an optimal design to evaluate its performance. Fig. 8 shows the design points with the minimum core energy-delay product (min-EDP) for different $L_{SPA}$'s and $L_{CON}$'s at $V_{DD}$ = 0.6 V and a fixed $L_{GATE}$ = 10 nm. The optimal design for min-EDP in this case happens at $L_{SPA}$ = 8 nm and $L_{CON}$ = 10 nm. To stress the importance of taking interconnects into account during device optimization, an artificial interconnect technology with lower resistivity is created to compare against the Cu case, which is assumed to have a 0.1× of the BEOL metal resistivity (including wires and vias) of Cu by modifying a parameter in the Steinhögl model. As shown in Fig. 8, the optimal device design shifts toward the left as the interconnect resistances ($R_{ITC}$) decrease because $R_{CON}$ then becomes more dominant such that a shorter $L_{SPA}$ and thus longer $L_{CON}$ provides more benefits.

### C. *Interconnect Analysis*

Evaluation of different wire materials other than Cu has been studied in [35]. In this section, wire distributions in the cores are analyzed to shed light on the complexity of interconnects. Interconnects in VLSI circuits/systems become increasingly complex with technology scaling due to increasing interconnect RC and more stringent lithography rules. Many design techniques (e.g. buffer insertion and gate sizing) are employed in modern place-and-route (P&R) EDA tools to place and connect standard cells on a die as densely as possible while meeting the timing requirements [41-42]. Analyses of wire distributions in the cores after timing closure for different $f_{TAR}$'s and $V_{DD}$'s are shown in Fig. 9. The average wire length normalized to CGP of each metal layer between two connected standard cells on the top 20 critical paths is shown in Fig. 9a. The low-level metal layers are much shorter than the high-level metal layers as the P&R tool manages to avoid using more resistive layers to transmit signals over long distances. Meanwhile, design-rule violations and routing congestions need be minimized, which makes the optimization problem even more complex. Average net length over an entire core and number of buffers (and inverters) vs. $f_{ACH}$ for different $V_{DD}$'s are shown in Fig. 9b and Fig. 9c. To achieve higher $f_{ACH}$'s for a fixed $V_{DD}$, the average net length becomes shorter as more buffers are inserted to break long wires. Moreover, the sizes of buffers also become larger to deliver more drive current to meet the timing requirement as shown in Fig. 9d. To quantify the impact of interconnect RC on the core speed, Fig. 9e and 9f show the contributions of $R_{ITC}$ and interconnect capacitance ($C_{ITC}$) to the critical-path delay ($t_{CP}$). The contribution is measured as $1 - t_{CP-R(C)}/t_{CP}$, where $t_{CP-R(C)}$ is the critical-path delay after zeroing out $R_{ITC}$ ($C_{ITC}$) in the extracted netlists of the critical path. As $f_{ACH}$ increases for a fixed $V_{DD}$, the impact of $R_{ITC}$ increases while the impact of $C_{ITC}$ decreases, because wider logic gates are required to deliver more drive current (i.e. lower cell-level resistance but higher capacitance). In general, $R_{ITC}$ and $C_{ITC}$ contribute to ~25% and ~40% of $t_{CP}$, respectively, across different $V_{DD}$'s. All in all, interconnects in a VLSI system are complex because various design techniques are used by the design tools to balance the interconnect RC against the transistor RC. Therefore, evaluation of the system-level performance without factoring in P&R optimization will be incomplete.

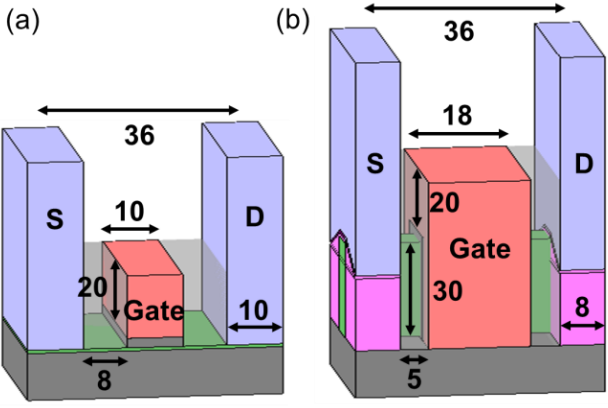

Fig. 10. Device structures optimized for minimum core energy-delay product. (a) Planar n-MoS$_2$/p-BP FET (b) Si FinFET. All numbers are in the unit of nm.

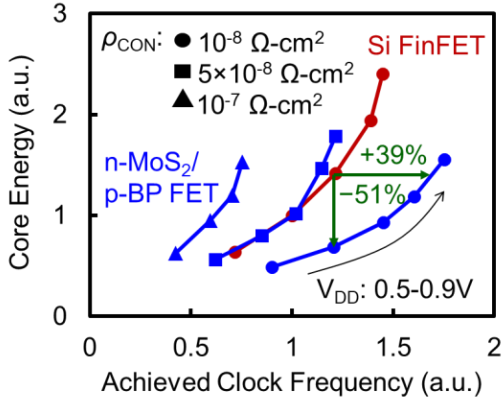

Fig. 11. Pareto optimal energy-frequency curves of the projected n-MoS$_2$/p-BP FET and the Si FinFET for different specific contact resistivities ($\rho_{CON}$).

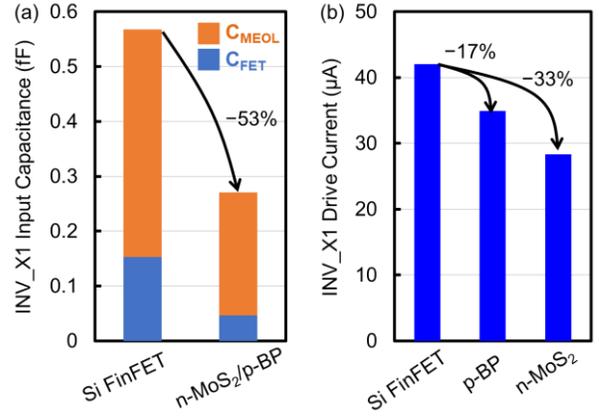

Fig. 12. (a) Input pin capacitance and (b) drive current (i.e. $V_{GS} = V_{DS} = V_{DD} = 0.6$ V) the INV_X1 standard cell for the projected Si FinFET and n-MoS$_2$/p-BP FET. $C_{MEOL}$ is the cell-level MEOL parasitic capacitance (see Fig. 3) and $C_{FET}$ is the transistor-level gate capacitance. For the Si FinFET case, there are 3 fins in the n and p active regions, respectively.

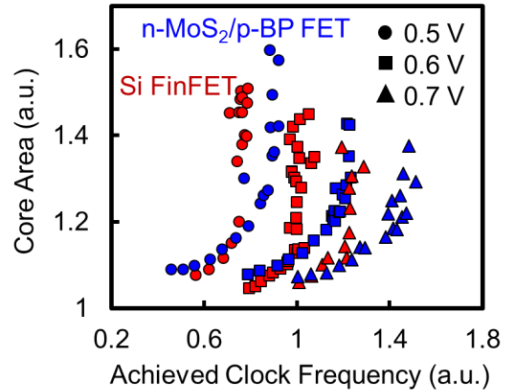

Fig. 13. Core area vs. clock frequency of implemented with the n-MoS$_2$/p-BP FET and the Si FinFET for different $V_{DD}$'s.

### D. n-MoS$_2$/p-BP Planer FETs vs. Si FinFET

n-MoS$_2$/p-BP FETs are compared against a projected 5-nm Si FinFET using DISPEL to study their potentials and challenges. While device performance benchmarking is itself an important topic requiring careful analysis [45-46], the aim here is not to arrive at a definitive assessment of a technology, but rather to demonstrate the importance of considering both parasitic and intrinsic parts as well as transistors and interconnects in the performance evaluation.

The VS model for the Si FinFET is fitted to the 14-nm data [43] to extract $\mu$ and $v$ and capture the I-V profile. Then the $\mu$ and $v$ are scaled up by 1.1× to match the projected intrinsic current assuming ballistic carrier transport [44]. The key VS model parameters and more details are summarized in

Appendix. The optimized device structure for Si FinFET following the same methodology described in Section III.B is shown in Fig. 10. The Pareto optimal E-f curves are compared between the theoretical n-MoS$_2$/p-BP FET and Si FinFET in Fig. 11. With the same assumed $\rho_{CON}$ of $10^{-8}$ Ω-cm$^2$, the MoS$_2$/BP FETs provide 51% lower iso-frequency energy consumption, or 39% faster iso-energy frequency compared to the Si FinFET. The superior energy efficiency is attributed to the much lower cell-level capacitance as shown in Fig. 12. Input capacitance of the INV_X1 cell based on the MoS$_2$/BP FETs is 53% lower than the Si-FinFET counterpart. 60% of the MoS$_2$/BP capacitance benefit comes from the reduced $C_{G2C}$ due to the longer $L_{SPA}$ (thanks to the better $L_{GATE}$ scalability) and its planar structure (as it doesn't have the gate-to-epi capacitance as illustrated in Fig. 3b), which highlights the importance of device and standard cell co-optimization to reduce parasitic RC. However, metal-to-2D-material contact resistances still limit the performance of 2D FETs to date [39]. As shown in Fig. 11, for the MoS$_2$/BP FETs to achieve comparable performance of the Si FinFET (with an assumed metal-to-Si $\rho_{CON}$ of $10^{-8}$ Ω-cm$^2$), a $\rho_{CON} < 5×10^{-8}$ Ω-cm$^2$ is required, which is ~6× lower than the $\rho_{CON}$ for monolayer MoS$_2$-to-metal contacts obtained from experiments [39]. Comparison of core area vs. frequency between the two transistor technologies is shown in Fig. 13.

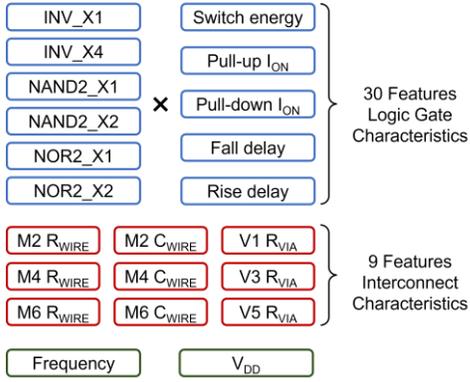

Fig. 14. 41 features are selected as the input to the neural-network model, including 30 features to characterize the logic performance (6 logic gates × 5 quantities measured from the chain of logic) and 9 features to characterize the interconnect performance.

Despite the lower drive current of the planar $MoS_2$/BP FETs, the core area is not necessarily larger than the Si FinFET for the same $f_{ACH}$ because of its lower cell capacitance. However, in the low frequency regime where the cell sizes are smaller in general (see Fig. 9d), the fixed capacitive loads at the core I/O pins become more important. Therefore, the $MoS_2$/BP FETs require more area (by ~3%) to drive the I/O pins and wire loads compared to the Si-FinFET counterpart.

To conclude this section, one of the key advantages of transistors based on one-/two-dimensional channel materials over bulk material-based transistors (e.g. Si FinFETs) is their low cell-level parasitic RC thanks to the superior electrostatic gate control, which enables further scaling of $L_{GATE}$ without the need to grow in the height of the device structure and gives more room for $L_{SPA}$ and $L_{CON}$. Therefore, it is important to factor in the cell-level capacitance and optimize the device structure and SC layout during technology assessment.

## IV. NEURAL-NETWORK PERFORMANCE PREDICTOR

While DISPEL provides accurate evaluation of system-level performance compared to the empirical models [11-14], the run time is much longer, ranging from hours to days, depending on the design complexity and constraints, which is not ideal for early technology assessment or design space exploration. Therefore, we train machine-learning (ML) models to predict system-level performance with technology-level parameters as the inputs. Most of the existing empirical models for system-level performance prediction rely on a set of explicit equations and empirical parameters (e.g. Rent's exponent, logic depth, fan-out), which require calibration to accurate data generated by full physical design flow. As discussed in section III.C, wiring optimization is such a complex process that it is very challenging for any explicit analytical models to predict the results accurately. In this regard, ML appear to be a reasonable approach since they are good at discovering intricate structure in large data sets. As DISPEL is highly automated, generating data for a variety of technology parameters becomes straightforward, making it feasible to train ML models.

To prepare training and testing data sets, we repeat the process introduced in Section III.A for a variety of combinations of technology parameters (e.g. different $\mu$'s and

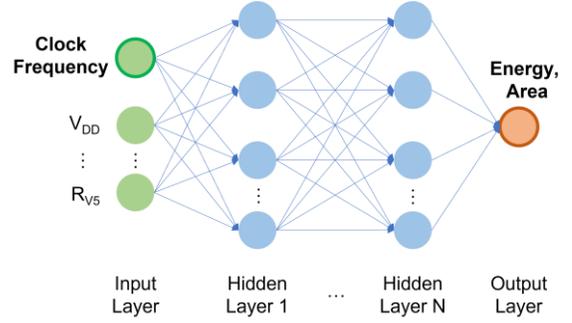

Fig. 15. A fully connected neuron network as the system performance predictor. Input: a vector of technology parameters and achieved clock frequency; output: core energy or area.

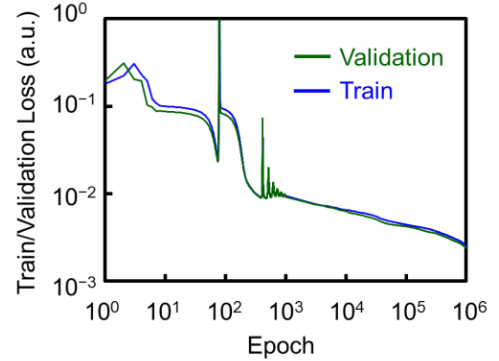

Fig. 16. A representative training and validation losses vs. epochs showing no signs of overfitting.

$v$'s in the FET models, $\rho_{CON}$'s, dielectric constants of gate spacers, BEOL metal resistivity, etc.), while the system architecture and design rules remain unchanged. Only the data points sitting on the Pareto-optimal curves for a fixed $V_{DD}$ are selected as they represent the optimal designs. The resulting data set has 2,763 samples. Each combination of technology parameters is then transformed into a vector of numbers, or *features*, as the inputs to the ML models, as illustrated in Fig. 14. The input features include: (i) 30 features characterizing the performance of 6 different logic gates. For each logic gate, 5 characteristics are derived from a simple fan-out-of-3 (FO3) circuit [62], including pull-up and pull-down drive currents ($I_{ON}$), falling and rising delays, and average switching energy consumption; (ii) 9 features characterizing the interconnect technology. Only M2, M4, and M6 are selected to represent the BEOL metal stacks because the physical properties (dimensions, resistivity) of M3/M5 are identical with M2/M4, and M1 is not used for routing; for vias, only V1, V3, and V5 are selected for the same reason, and only the resistances are considered because via capacitances are negligible; (iii) the last two features are $V_{DD}$ and $f_{ACH}$. The model output is either the core energy or area. Next, a fully connected NN model as illustrated in Fig. 15 with the following designs are trained using TensorFlow [49]: (i) the activation function of each neuron is a Softplus function [50], i.e. $f(x) = \ln(1+e^x)$; (ii) mean squared error is used as the regression's loss function; (iii) the Adam algorithm [51] is used as the optimizer. Before training, input vectors are rescaled to a range of [−1, 1], and the neuron weights are initialized by the Xavier initialization [52] for training stability. The entire data set is split into training and test data sets. The training set is further divided into 80/20 ratios for training and validation, respectively. Hyper-parameters such as learning rates, L2 regularization, and the architecture of NNs are experimented to minimize the validation error. We found empirically that a 2(-hidden)-layer NN with 40 neurons on the first hidden layer and 20 neurons on the second layer achieves the minimum losses (~4% test loss). Fig. 16 shows one representative training curve over one million epochs without any signs of overfitting.

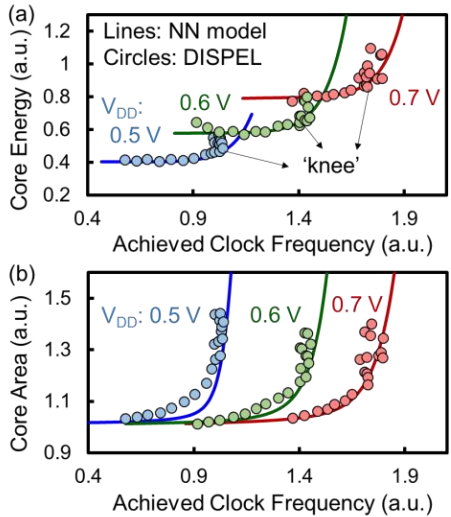

Fig. 17. Comparison of the core (a) energy and (b) area vs. frequency between the implementation results using DISPEL and the neural-network (NN) model predictions for n-MoS$_2$/p-BP FET+BEOL interconnect with 50% wire resistance and 25% lower capacitance.

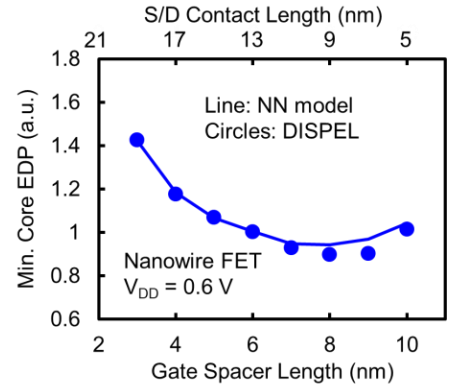

Fig. 18. Comparison of minimum core energy-delay product vs. gate spacer length between the implementation results using DISPEL and the neural-network (NN) model predictions for nanowire FETs with fixed CGP = 36 nm and $L_{GATE}$ = 11 nm at $V_{DD}$ = 0.6 V.

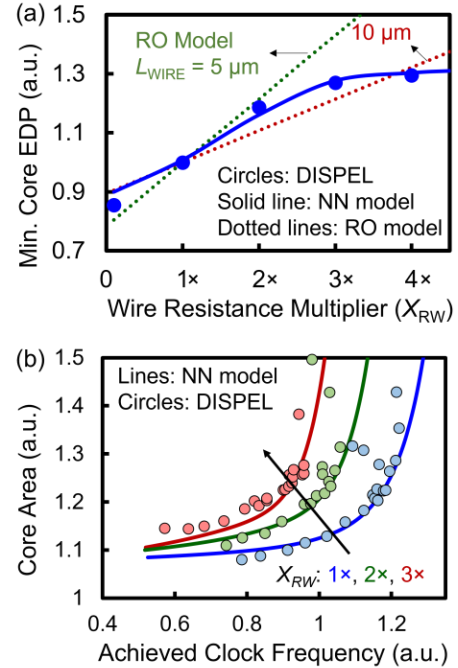

Fig. 19. (a) Comparison of minimum energy-delay product vs. wire resistance multiplier ($X_{RW}$) between the DISPEL implementation results and neural-network (NN) predictions. The dotted lines are predictions of ring-oscillator (RO) models with different lengths of the wire loads. (b) Core area vs. frequency for different $X_{RW}$'s.

While high accuracy is necessary for a good ML model, it is also important to verify its physical robustness. To this end, three test sets are created. Each test is composed of a combination of transistor and interconnect technologies that the model has never seen in the training data set. The first test is to test if the model can capture the relations between the key performance metrics — core energy consumptions, areas, and clock frequencies. The test data is based on a hypothetical interconnect technology with a BEOL wire resistivity ($R_{WIRE}$) 50% lower than the baseline case of Cu and a 25% lower interlayer dielectric constant (i.e. lower wire capacitances). Fig. 17 shows the predicted Pareto optimal curves of energy and area vs. frequency compared against the results generated by DISPEL. To generate the predictions, the input clock frequency to the NN is swept from low to high while the rest of the input features are fixed. The predicted energy consumptions and core areas increase smoothly and monotonically with increasing frequencies at an accelerating rate, which is a result of the shallow NN with just two hidden layers and is more physically meaningful over the results of other models with higher complexity. In-depth analysis of the model is discussed in Section V. Note that the deviation of the predictions from the implementation results (through DISPEL) is larger in the high frequency regime. To achieve higher frequencies, more larger logic gates are needed. Beyond a certain point, the capacitances of the logic gates on the critical paths become so dominant that making the gates bigger does not increase the frequency anymore, and thus the implementation results become less predictable in the high frequency regime. Nonetheless, it is the region around the 'knees' of the curves (illustrated in Fig. 17) that matter the most because it is where the most efficient designs (e.g. min-EDP) reside in.

The second test is to test if the model can be used to explore the optimal device structure for a given performance metric. The test data is based on a stacked nanowire (NW) FET model fitted to numerical simulations [53] and is not present in the training data set. Performance analysis of NW FETs has been studied elsewhere [40,54] and is not the focus in this paper so the NW FETs here should be viewed as yet another new transistor technology to be explored. Fig. 18 shows the predicted core min-EDP at $V_{DD}$ = 0.6 V compared against the implementation results for different $L_{SPA}$'s and $L_{CON}$'s (see Fig.

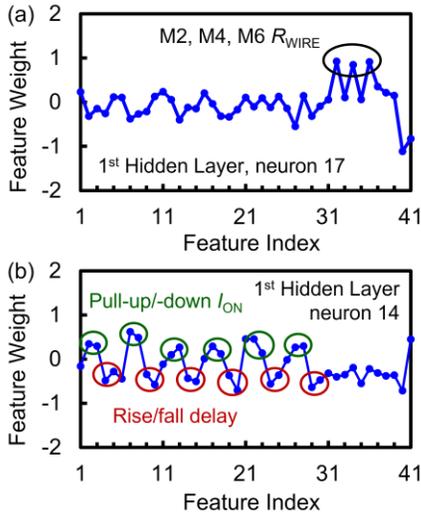

Fig. 20. Weights of the 41 input features to two of the neurons on the first hidden layer: (a) the 17th neuron reactive to BEOL wire resistivity. (b) The 14th neuron reactive to logic gate delays and drive currents ($I_{ON}$).

3) with fixed $L_{GATE}$ and CGP. While the predictions of the min-EDP are slightly off at the long $L_{SPA}$ regime, the predicted optimal $L_{SPA}$ and $L_{CON}$ are reasonably close to the DISPEL results.

The third test is to test if impact of $R_{WIRE}$ on the core performance is captured in the model, as $R_{WIRE}$ is expected to increase rapidly with dimensional scaling. The test data is created by artificially multiplying $R_{WIRE}$ of M2 to M6 by a factor of $X_{RW}$ (from 0.1 to 4) in the ITF file in the DISPEL workflow. The predicted min-EDPs of the core vs. $X_{RW}$ are compared with the implementation results in Fig. 19a. The min-EDP grows sub-linearly with increasing $X_{RW}$ as the P&R tool manages to reduce the impact of wire resistances by, for instance, inserting more buffers. Consequently, the core areas are also increased with increasing $X_{RW}$ as shown in Fig. 19b. The NN model is able to capture the nonlinear relation between EDP and $X_{RW}$, whereas a ring-oscillator model with a fixed-length wire load would predict a linear increase in EDP as $X_{RW}$ increases (as shown by the dotted lines in Fig. 19a).

## V. ANALYSIS OF NEURAL-NETWORK MODEL

In-depth analysis of the 2(-hidden)-layer NN model introduced in Section IV is presented in this section. The first hidden layer (HL$_1$) has 40 neurons. Each neuron has 41 weights corresponding to the 41 input features (see Fig. 15). The weights reflect the sensitivity of each neuron to different features. For instance, Fig. 20 shows that the 17th neuron is particularly reactive to $R_{WIRE}$, while the 14th neuron is reactive to the logic gate characteristics. The weights corresponding to the logic gate $I_{ON}$ and delay features have opposite polarities, which matches the intuition because large $I_{ON}$'s are preferred for performance whereas large delays are unfavorable. The second hidden layer (HL$_2$) has 20 neurons and start to become too opaque to draw insights from their weights. Nonetheless, as shown in Fig. 21, it is found that only one neuron transitions from an inactive state (i.e. the output is close to 0) to an active state (i.e. the output is >> 0) as the input frequency increases.

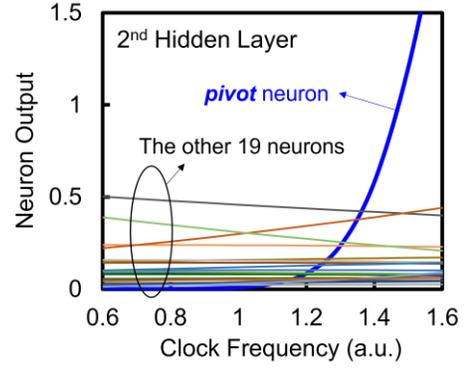

Fig. 21. Outputs of the 20 neurons on the second hidden layer vs. input frequency. The *pivot* neuron is the only neuron that transitions from being inactive to active across the frequency range while the other neurons always stay either inactive or active.

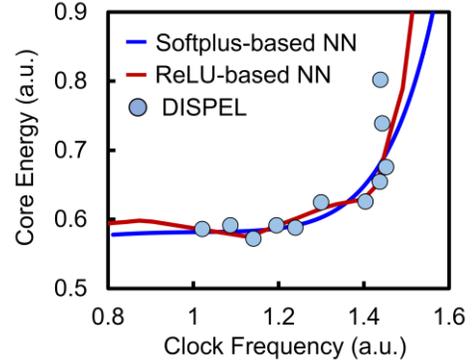

Fig. 22. Comparison of the predicted energy-frequency curves between two neural-network models with different activation functions: the Softplus vs. Rectifier functions.

The other neurons always stay either inactive or active across the frequency range. It is the transition of this *pivot* neuron from being inactive to active that leads to the "hockey-stick" shape of the energy (and area) vs. frequency relations, and the smooth transition is a characteristic of the Softplus function. The effects of other neurons are mainly to shift the E-f (and A-f) curves in the vertical and/or lateral directions. For example, an increase in $R_{WIRE}$ raises the output of HL$_1$'s 17th neuron, which activates the pivot neuron on HL$_2$ earlier and shifts the curves toward the left, i.e. higher energy and larger area for the same frequency. Interpretation of what representations has a NN learned is still an active research area [55]. To see the effect of using the Softplus function, another NN with Rectified Linear Units (ReLUs) [56] (i.e. $f(x) = \max(0, x)$) as the activation functions is trained and the result is shown in Fig. 22 compared against the predictions of the Softplus-based NN. The zigzag pattern is apparent in the output of the ReLU-based NN because the output of a ReLU is a piecewise linear function, and overfitting becomes more likely to happen. Similar outcomes of overfitting are also observed in other ML models such as NN with more hidden layers or random forest regression models. The 2-hidden-layer NN with Softplus activation functions is found to be the optimal model architecture that gives both accurate predictions as well as smooth outputs which reduces the chance of overfitting.

## VI. DISCUSSION AND OUTLOOK

In this section, limitations of the DISPEL workflow and the ML-based performance predictors as well as possible ways to improve them are discussed. While only nominal cases at the typical corner were analyzed, PVT variations can be easily incorporated in DISPEL through multi-corner multi-mode analysis [57]. The key challenge is to create models that capture the process variations in the new transistor and interconnect technologies, which is a non-trivial task because new technologies are often not mature enough to provide sufficient amounts of data at different corners. Similarly, while integrating memory instances (e.g. SRAM) into DISPEL is possible, creating memory models and compilers [58] that accurately capture PVT variations to ensure adequate margins for millions of memory cells requires significant amount of work.

The main idea of the DISPEL workflow is to streamline the process from end to end to provide a holistic view of the impact of CMOS technologies on the system-level performance. Performance evaluation across different technology nodes can provide insights into the benefits of further dimensional downscaling and design guidance. However, several parts are skipped in this paper for simplicity, including the design constraints of different lithography technologies and floor plan optimization. For more rigorous and realistic assessment, proper design rules and floor plan optimization need to be taken into account, which by themselves are also big research topics.

Thanks to the highly automated workflow of DISPEL, large amount of data can be generated to leverage the power of ML algorithms to discover the dependencies of system-level performance on technology-level design parameters. The NN model presented in this paper is an attempt the performance of a specific system (i.e. the 32-bit processor core) at a particular node (i.e. the projected 5-nm node). To generalize the method to different system architectures and/or technology nodes, the input features must be modified to encapsulate the architectural information. The Rent's exponent and constants is an example to abstract a certain type of system in a few numbers [11]. To account for dimensional downscaling in a more general sense, CGP and interconnect dimensions need to be treated as free input variables. And just like all the ML problems, feature engineering can make a big difference. One can choose device- or process-level parameters such as $L_{GATE}$, CGP, $\mu$, or $\rho_{CON}$ rather than the logic-level features. In any case, a large amount of data is required, which calls for a highly integrated and automated design flow across the boundaries between devices, interconnects, circuits, and systems like DISPEL.

## VII. CONCLUSION

The Device-to-System Performance EvaLuation (DISPEL) platform presented in this paper provides a framework for assessment of new transistor and interconnect technologies from the standpoint of system-level performance through a highly integrated workflow from transistor and interconnect modeling to physical design flow. Full-chip placement and routing enables accurate evaluation of the system performance that simple benchmark circuits cannot offer. Using DISPEL, we demonstrate how device structures can be optimized to reduce the impact of parasitic RC on the performance of a 32-bit processor core at the projected 5-nm node and provide a more accurate view of the advantages of 2D-channel-material FETs and their challenges. Large amount of data generated by the highly automated DISPEL workflow is used to train neural-network models to predict the performance of the processor core. A two-hidden-layer neuron network with Softplus activation functions is found to achieve the most accurate and physically favorable results. As technology scaling becomes ever more challenging, highly integrated and automated design flows across the boundaries between devices, interconnects, circuits, and systems like DISPEL can facilitate technology development and provide design guidance in the early stage.

TABLE II. KEY VIRTUAL SOURCE MODEL PARAMETERS

| Parameter | Theoretical n-MoS$_2$ FET | Theoretical p-BP FET | Projected Si FinFET |
|---|---|---|---|
| $v$ ($10^7$ cm/s) | 1.17 | 1.7 | 0.97 |
| $\mu$ (cm$^2$/V-s) | 200 | 350 | 253 |
| $L_{GATE}$ (nm) | 10 | 10 | 18 |
| $C_{INV}$ ($\mu$F/cm$^2$) | 4.36 | 4.26 | 3.14 |
| Fin Width /Height/Pitch (nm) | N/A | N/A | 5/30/21 |
| SS (mV/dec) | 70 | | |

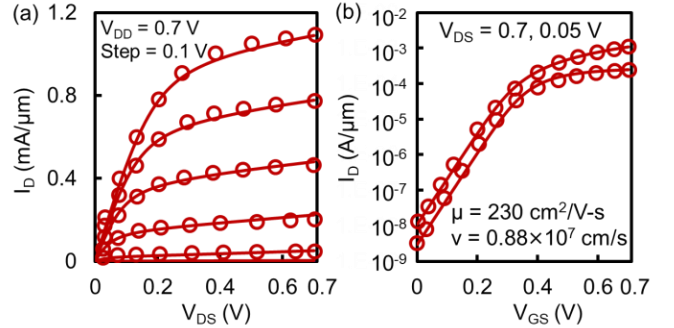

Fig. 23. Virtual Source model (lines) fitted to the 14-nm Si FinFET data [43] (circles) to extract carrier mobility ($\mu$) and velocity ($v$). (a) $I_D$-$V_{DS}$ and (b) $I_D$-$V_{GS}$. Effective thickness of the gate oxide is assumed to be 1.2 nm.

## APPENDIX

Key parameters of the VS models for the theoretical n-type MoS$_2$ FET, p-type BP FET, and the projected Si FinFET at the 5-nm node are listed in in Table II. For the n-MoS$_2$ and p-BP FETs, $\mu$ and $v$ are extracted from the current-/capacitance-voltage characteristics of physics-based numerical simulations [18, 19, 48] and should be viewed as theoretical predictions; for the Si FinFET, $\mu$ and $v$ are extracted from the 14-nm FinFET experimental data [43] (Fig. 23) and scaled up by 1.1× to match the projection assuming ballistic transport [44]. $L_{GATE}$ is set at the value such that the subthreshold slope is 70 mV/dec based on numerical simulations. The inversion gate-to-channel capacitance ($C_{INV}$) is derived from $C_{OX} \cdot C_Q / (C_{OX} + C_Q)$, where $C_{OX} = \varepsilon_{SiO2}/EOT$, $C_Q$ is the quantum capacitance, $\varepsilon_{SiO2}$ is the permittivity of SiO$_2$, and EOT = 0.7 nm.

ACKNOWLEDGMENT

This work was supported in part through the NCN-NEEDS program, which was funded by the National Science Foundation, contract 1227020-EEC, and by the Semiconductor Research Corporation, and through Systems on Nanoscale Information fabriCs (SONIC), and Function Accelerated NanoMaterials Engineering (FAME), two of the six SRC STARnet Centers, sponsored by MARCO and DARPA, as well as the member companies of the Stanford SystemX Alliance and the Initiative for Nanoscale Materials and Processes (INMP) at Stanford University.